\documentclass[11pt, prd,preprintnumbers,amsmath,amssymb, superscriptaddress]{revtex4}

\pdfoutput=1
\usepackage{latexsym}
\usepackage{amssymb}
\usepackage{epsfig,amsmath,graphics}
\usepackage{epstopdf}
\usepackage{verbatim}
\usepackage[hypertex]{hyperref}
\usepackage{graphicx}
\usepackage{subfigure}

\newcommand{\OO}{\mathcal{O}}

\newcommand{\GeV}{\text{GeV}}
\newcommand{\TeV}{\text{TeV}}

\newcommand{\Qn}{Q_4}
\newcommand{\QnBar}{\overline{Q}_4}
\newcommand{\Un}{U_4}
\newcommand{\UnBar}{\overline{U}_4}
\newcommand{\Dn}{D_4}
\newcommand{\DnBar}{\overline{D}_4}
\newcommand{\Ln}{L_4}
\newcommand{\LnBar}{\overline{L}_4}
\newcommand{\En}{E_4}
\newcommand{\EnBar}{\overline{E}_4}
\newcommand{\Hu}{H_u}

\newcommand{\muQ}{\mu_Q}
\newcommand{\muU}{\mu_U}
\newcommand{\muD}{\mu_D}
\newcommand{\muL}{\mu_L}
\newcommand{\muE}{\mu_E}
\newcommand{\munew}{\mu_4}

\newcommand{\OrderOne}{\mathcal{O}\left(1\right)}

\newcommand{\Mgut}{M_\text{GUT}}

\def\r{\right)}
\def\l{\left(}

\begin{document}


\title{A Little Solution to the Little Hierarchy Problem: A Vector-like Generation}

\author{Peter W. Graham}
\affiliation{Department of Physics, Stanford University, Stanford, California 94305}

\author{Ahmed Ismail}
\affiliation{Department of Physics, Stanford University, Stanford, California 94305}

\author{Surjeet Rajendran}
\affiliation{Center for Theoretical Physics, Laboratory for Nuclear Science and Department of Physics, Massachusetts Institute of Technology, Cambridge, MA 02139, USA}
\affiliation{SLAC National Accelerator Laboratory, Stanford University, Menlo Park, California 94025}
\affiliation{Department of Physics, Stanford University, Stanford, California 94305}

\author{Prashant Saraswat}
\affiliation{Department of Physics, Stanford University, Stanford, California 94305}

\preprint{MIT-CTP 4078}

\date{\today}

\begin{abstract}
We present a simple solution to the little hierarchy problem in the MSSM: a vector-like fourth generation.  With $\OO(1)$ Yukawa couplings for the new quarks, the Higgs mass can naturally be above 114 GeV.  Unlike a chiral fourth generation, a vector-like generation can solve the little hierarchy problem while remaining consistent with precision electroweak and direct production constraints, and maintaining the success of the grand unified framework.  The new quarks are predicted to lie between $\sim 300 - 600$ GeV and will thus be discovered or ruled out at the LHC.  This scenario suggests exploration of several novel collider signatures.
\end{abstract}

\maketitle

\section{Introduction}
\label{Sec:Intro}

The hierarchy problem has for years been taken as a strong motivation for theories of physics beyond the Standard Model (SM).  The Minimal Supersymmetric Standard Model (MSSM) is one of the most attractive ideas for solving this problem as it naturally gives gauge coupling unification and a dark matter candidate.  However the MSSM predicts a light Higgs boson, near the Z mass, while LEP placed a lower limit on the Higgs mass of 114 GeV.  To satisfy the LEP bound, the stop quark must be taken to be $\sim 1 ~\TeV$ so that radiative corrections from the top quark increase the Higgs mass sufficiently.  Thus supersymmetry (SUSY) must be broken above the weak scale, recreating a fine-tuning of $\sim 1 \%$ or worse in the soft SUSY-breaking parameters in order to reproduce the observed value of the weak scale.  This is how the little hierarchy problem appears in the context of the MSSM \cite{Barbieri:2000gf, Giudice:2006sn, Raby:2007yv, Dermisek:2009si}.  In this paper we point out that a vector-like fourth generation can solve this problem by adding extra radiative corrections to the Higgs mass.

This solution is straightforward, relying mostly on having new quarks, and is thus predictive.  In order to remove the fine-tuning and avoid current experimental constraints there must be at least one new colored particle with mass between roughly 300 GeV and 600 GeV, easily discoverable at the LHC.  As we will show, this solves the little hierarchy problem while naturally preserving the success of unification.  Alternative solutions to the little hierarchy problem in the MSSM either involve large couplings which spoil unification or require new gauge or global symmetries, a very low messenger scale or a carefully chosen set of soft SUSY-breaking parameters \cite{Choi:2005hd, Kitano:2005wc, Chacko:2005ra, Ellis:1988er, Espinosa:1998re, Batra:2003nj, Maloney:2004rc, Casas:2003jx, Brignole:2003cm, Harnik:2003rs, Chang:2004db, Delgado:2005fq, Birkedal:2004zx, Babu:2004xg, Kim:2006mb, Choi:2006xb, Dutta:2007az, Abe:2007kf, Dutta:2007xr, Abe:2007je, Dine:2007xi, Dermisek:2009si, Gogoladze:2009bd, Dermisek:2006ey, Kikuchi:2008ws, Bellazzini:2009ix}.  Other solutions involve extensions of the Higgs sector to create unusual decays of the Higgs in order to avoid LEP bounds \cite{Dermisek:2005ar, Bellazzini:2009xt}.  Twin and little Higgs theories have also been proposed to solve the little hierarchy problem \cite{Chacko:2005pe, Burdman:2006tz, Chang:2006ra, Falkowski:2006qq, ArkaniHamed:2002qy, ArkaniHamed:2001nc} by extending the symmetries of the standard model to a larger structure with a collective breaking pattern, though these theories only push the cutoff up to $\sim 10$ TeV \cite{Grzadkowski:2009mj}.  A chiral fourth generation has been considered before but does not solve the little hierarchy problem and runs into difficulty with the large Yukawa couplings necessary to avoid experimental constraints leading to low scale Landau poles \cite{Kribs:2007nz, Fok:2008yg, Richard:2008ye, Ibrahim:2008gg, Murdock:2008rx, Liu:2009cc, Litsey:2009rp, Bobrowski:2009ng}.  A vector-like generation has been proposed \cite{Liu:2009cc, Barger:2006fm} but its possible use in solving the Little Hierarchy problem was only appreciated in \cite{Babu:2008ge}.  We discuss in Section \ref{Sec:Constraints} why we believe the problem is more fully ameliorated than was claimed in \cite{Babu:2008ge}.

In Section \ref{Sec:TheModel} the model is presented.  Section \ref{Sec:RGE} presents the renormalization group analysis.  Section \ref{Sec:Spectrum} presents the physical masses of the new particles.  In Section \ref{Sec:Constraints} we calculate the experimental constraints on our model from direct collider production and precision electroweak observables.  In Section \ref{Sec:HiggsMass} we evaluate the Higgs mass.  In Section \ref{Sec:Pheno} we discuss collider signatures of this model.

\section{The Model}
\label{Sec:TheModel}

We add a full vector-like generation
to the MSSM with the following Yukawa interactions
\begin{equation}
\mathcal{W} \supset y_4 \, \Qn \Un \Hu  \, + \, z_4 \, \QnBar \DnBar \Hu
\label{Eqn:NewYukawas}
\end{equation}
and mass terms
\begin{eqnarray}
\mathcal{W} & \supset & \muQ \Qn \QnBar \, + \, \muU \Un \UnBar \, + \, \muD \Dn \DnBar \, + \, \muL \Ln \LnBar \, + \, \muE \En \EnBar 
\label{Eqn:NewMasses}
\end{eqnarray}
 in the superpotential.  The subscript $4$ denotes the new generation. In equations \eqref{Eqn:NewYukawas} and \eqref{Eqn:NewMasses} and the  rest of the paper, we use the familiar notation of the MSSM \cite{Martin:1997ns}. The superpotential \eqref{Eqn:NewYukawas} implicitly assumes a discrete parity under which the new matter is charged. This parity forbids mixing between the new generation and the standard model. This parity does not affect the Higgs mass in this model but has other interesting phenomenological consequences that are discussed in section \ref{Sec:Parity}. It is also possible to write the model without this parity in which case the first term in Eqn.~\eqref{Eqn:NewYukawas} is extended to a full 4x4 Yukawa matrix allowing mixing between all the generations.  These mixings, if present,  have to be small from FCNC limits \cite{Bobrowski:2009ng, Chanowitz:2009mz} and we will assume this to be the case.

 Upon SUSY breaking, the terms in \eqref{Eqn:NewYukawas} contribute to the Higgs quartic. Including the contribution from the top Yukawa $y_3$, the Higgs mass $m_h$ in this model is roughly given by
\begin{eqnarray}
m_h^2  &\sim& M_{z}^2 \cos^{2} 2\beta  \, + \, \l \frac{3}{2 \pi^2}\r v^2 \sin^{4} \beta  \l y_3^4 \log \frac{m_{\tilde{t}}}{m_t}  \, +  \, y_4^4 \log \frac{m_{\widetilde{\Qn}}}{m_{\Qn}}  + \, z_4^4 \log \frac{m_{\widetilde{\QnBar}}}{m_{\QnBar}} \r  
\label{Eqn:Estimate}
\end{eqnarray}
where $v \sim 174$ GeV is the electroweak symmetry breaking vev. The contributions from the new Yukawa couplings add linearly to $m_h^2$ and so can increase $m_h^2$ more effectively than the usual logarithmic contribution from raising the stop masses. As a result, this model can be compatible with the LEP limit on the Higgs mass  with smaller soft scalar masses, and is significantly less  tuned.  We calculate the Higgs mass more precisely in Section \ref{Sec:HiggsMass}.

For example, with  $y_4 \approxeq z_4 \approxeq y_3$,  the size of the logarithmic corrections in \eqref{Eqn:Estimate} is roughly three times that of the top sector alone. In this case,  a Higgs mass $\sim 114$ GeV  can be obtained with soft masses $\sim 300$ GeV (taking $\tan \beta \sim 5$ and the vector masses $\muQ, \muU, \muD \sim 300$ GeV).  For similar parameters, in the MSSM, the stop has to be $\gtrsim 1.1$ TeV  in order that $m_h > 114$ GeV \cite{Martin:1997ns}. Since the Higgs vev is quadratically sensitive to the soft scalar masses, we expect the tuning in our model to be alleviated by a factor of $\sim \l \frac{\text{1.1 TeV}}{\text{300 GeV}}\r^2 \sim \OO\l 10\r$.

We first make some qualitative remarks about the parameter space of the model. The corrections to $m^2_h$ from the new generation scale as the fourth power of the couplings $y_4$ and $z_4$  (see equation \eqref{Eqn:Estimate}). If these couplings are much smaller than the top Yukawa, their effects become quickly subdominant. Moreover,  these Yukawas renormalize each other and the top Yukawa and can lead to UV Landau poles. Motivated by gauge coupling unification, we impose the requirement that these Landau poles lie beyond the GUT scale. This sets an upper bound on the low energy values of $y_4$ and $z_4$. Since $y_3$ is close to its fixed point, we expect this bound to lie around the fixed point.  These two considerations lead us to expect $y_4$ and $z_4$ to lie in a technically natural, but narrow, range around $y_3$. 

The superpotential \eqref{Eqn:NewYukawas} can also contain Yukawa terms  between the Higgs sector and the leptonic components of the new generation ({\it e.g} $w_4 \LnBar \EnBar H_u$). These terms will also contribute to the Higgs quartic. However,  the color factor for these loops is a third of the color factor for the quark loops. Hence,  we expect these corrections  to be subdominant, unless the couplings are large. But, these leptonic Yukawas become non-perturbative more easily than the quark Yukawas since their one loop beta functions are unaffected by the strong coupling constant $g_3$ (see Section \ref{Sec:RGE}). This constrains these Yukawas to be be smaller than the corresponding quark Yukawas and hence they do not make significant corrections to the Higgs mass. In this paper, we assume that these Yukawas are small and ignore their effects on the phenomenology. 

The contributions to $m^2_h$ from the new vector-like generation is a function of SUSY breaking in that sector and is suppressed by $\sim \frac{ \widetilde{m}_{\Qn}^2}{m_{\Qn}^2}$. Here $\widetilde{m}_{\Qn}^2$, the soft mass, is the difference between the scalar and fermion mass squares respectively. These contributions are unsuppressed when  $\widetilde{m}_{\Qn}^2 \sim  m_{{\Qn}}^2$. Since $\widetilde{m}_{\Qn}^2$ contributes quadratically to the Higgs vev, the tuning in this model is minimized when $\widetilde{m}_{\Qn}^2 \sim  \l200 \text{ GeV}\r^2$. This leads us to expect the masses of the new generation to lie around $\sim 200$ GeV - a range easily accessible to the LHC.

\section{The Renormalization Group Analysis}
\label{Sec:RGE}

In this section, we study the renormalization group evolution of all the parameters. We identify the regions of the $y_4$-$z_4$ parameter space where the theory is free of Landau poles up to the GUT scale.  The addition of the new vector-like generation also affects the evolution of gauge couplings.  Since the new particles form complete $SU(5)$ multiplets, gauge coupling unification is preserved in this scenario. However, the extra matter fields do change the running of gaugino and soft scalar masses. 

The evolution of the gauge couplings $g_i$ are governed by the equations \cite{Martin:1997ns}
\begin{eqnarray}
\label{Eqn: gauge RGEs}
\frac{d}{dt}g_i & = & \frac{1}{16 \pi^2} b_i g_i^3
\end{eqnarray} 
With the particle content of this model  $\l b_1, \, b_2, \, b_3 \r = \l \frac{53}{5}, \, 5, \, 1 \r$, and the gauge couplings unify perturbatively at roughly $\sim 10^{16}$ GeV. The running of the Yukawa couplings are governed by 
\begin{eqnarray}
\frac{d}{dt} y_i & = & \frac{1}{16 \pi^2} y_i \, \l 6 \l y_3 y_3^{*} + y_4 y_4^{*} \r +  3 \,  z_4 z_4^{*}  - \l \frac{16}{3} g_3^2 + 3 g_2^2 + \frac{13}{15}g_1^2 \r  \r \\
\frac{d}{dt} z_4 & = & \frac{1}{16 \pi^2} z_4 \, \l 3 \l y_3 y_3^{*} + y_4 y_4^{*} \r +  6 \,  z_4 z_4^{*}  - \l \frac{16}{3} g_3^2 + 3 g_2^2 + \frac{7}{15}g_1^2 \r  \r 
\label{Eqn:BetaFunctions}
\end{eqnarray}
In writing the above, we have ignored contributions to these expressions from the Yukawa couplings in the down and lepton sector. This is reasonable in the regime of moderate $\tan \beta \lessapprox 10$ where the down and lepton Yukawas are small. Similarly, we have also ignored mixing terms between the vector-like generation and the standard model, since these Yukawas also have to be small to evade FCNC constraints. 

\begin{figure}
\begin{center}
\includegraphics[width = 3.0 in]{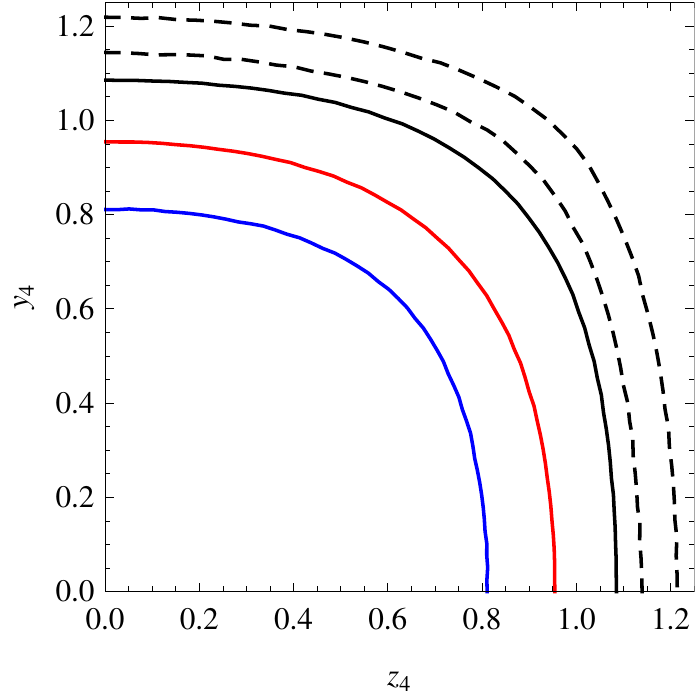}
\caption{ \label{Fig:yzplot} (Color online) The parameter space in the $z_4$-$y_4$ plane which solves the little hierarchy problem.  The dashed lines show the maximum values of $y_4$ and $z_4$ such that all couplings remain perturbative (do not hit a Landau pole) up to the GUT scale (lower line) or $10^9$ GeV (upper line).  The solid lines show the minimum values of $y_4$ and $z_4$ that bring the Higgs mass above 114 GeV.  From top to bottom they are $(\tan \beta, A) = $ $(5, 300 \, \GeV)$ (black), $(5, 350 \, \GeV)$ (red), $(10, 300 \, \GeV)$ (blue), where $A$ is the unified A-term value.  Here we have taken a unified vector mass $\mu_4 = \muQ = \muU = \muD = 320 ~ \GeV$, a unified soft mass $\widetilde{m}^2 =  \widetilde{m}_{Q}^2 = \widetilde{m}_{{U}}^2 =  \widetilde{m}_{{D}}^2 = \left( 350 ~\GeV \right)^2$, and the decoupling limit.}
\end{center}
\end{figure}

The running of these couplings is governed by the competition between the Yukawa and the gauge couplings. Since the gauge interactions themselves get stronger in the UV, in particular the strong coupling $g_3$ (see Eqn.~\ref{Eqn: gauge RGEs}), the model is able to accommodate low energy values of $y_4,\, z_4 \sim 0.9$ without any coupling hitting a Landau pole below the GUT scale, in order to preserve perturbative unification. We plot this parameter space in the $y_4$ - $z_4$ plane in Figure \ref{Fig:yzplot}.

The above perturbativity analysis was performed at the one loop level. Higher
loop contributions to the beta functions were included in the analysis of 
\cite{Martin:2009}. These additional contributions cause the gauge couplings to
become non-perturbative roughly around the unification scale $\sim 10^{16}$ GeV.
This suggests that in the presence of a vector-like generation, the physics at
the unification scale may be more compatible with orbifold GUT constructions
instead of simple 4D unification scenarios. We note that these orbifold
constructions offer several advantages over simple 4D unification scenarios,
including natural ways to incorporate doublet-triplet splitting and avoiding dimension 5
proton decay constraints.  It is also possible that the flavor structure of the
SM is generated at a scale below the GUT scale and thus the Yukawa couplings
need only remain perturbative up to that intermediate scale, which will expand
the available parameter space.  We show an example where flavor is generated
below $10^9 \, \GeV$ in Figure \ref{Fig:yzplot}.

The modified gauge couplings also affect the running of the soft gaugino and scalar masses. The weak scale soft scalar masses were computed (using \cite{Martin:1993zk}) after imposing the condition that the low energy gluino and electroweakino masses obey current bounds ($M_3 > 300$ GeV, and $M_1, \, M_2 > 100$ GeV) \cite{PDG}. With this constraint on the gaugino masses, we find that the typical size of soft scalar masses in this model is $\sim 400 \text{ GeV} + 50 \, \GeV  \l\log \l \frac{M_S}{10^9 \text{ GeV}}\r\r$, where $M_S$ is the scale at which the soft masses ($\sim 100$ GeV) are generated. A primordial SUSY breaking scale $M_S$ larger than  $10^9$ GeV will drag the soft scalar masses up and reintroduce tuning. SUSY breaking at scales $\lessapprox 10^9$ GeV are natural in many models of SUSY breaking {\it e.g.}~gauge mediation. These SUSY breaking models also address many of the other problems that plague the MSSM and they can naturally accommodate our framework. Note that if the soft scalar masses are universal for the various generations at the scale $M_S$ (as one might expect for example in gauge mediated SUSY breaking), then the running to the weak scale does not induce large variation between the generations. We will generally assume universal weak scale soft scalar masses for the third and fourth generations when determining the Higgs mass within this model.

The large Yukawa couplings and SUSY breaking gaugino masses also drive the generation of $A$ terms. These $A$ terms also make small, but significant corrections to the Higgs mass. As with the Yukawa couplings, when computing the Higgs mass we consider only the $H_u$ type $A$ terms: $\mathcal{L} \supset (-A^{ij}_{y} \tilde{Q}_i {\tilde{U}}_j^* H_u - A_{z} \tilde{\QnBar} \tilde{\DnBar^*} H_u)$.  The beta functions of $A$ terms consist of terms proportional to themselves and terms proportional to the corresponding Yukawa couplings. With the Yukawa couplings of interest to this paper, $A$ terms $\sim -\l 300  \text{ GeV} + 30 \, \GeV  \l\log \l \frac{M_S}{10^9 \text{ GeV}}\r\r\r$ are generated for the third and fourth generations even when they are zero at the SUSY breaking scale. For example, gauge mediation typically gives zero $A$ terms at the SUSY breaking scale. In the context of such theories, renormalization group evolution can lead to weak scale $A$ terms  $\sim -300  \text{ GeV} $ for the generations that have $O(1)$ Yukawa couplings, with much smaller values for the first two generations. In other SUSY breaking scenarios such as gravity mediation,  the $A$ terms are free parameters and can also easily be $\sim -300$ GeV. Using this input from the renormalization group flow, in this paper, we generally take $A$ terms $\sim -300$ GeV for the third and fourth generations while computing the corrections to the Higgs mass.  

The vector masses run proportional to themselves, and their running is affected by both the Yukawa couplings ($y_4$ and $z_4$) and the gauge couplings $g_i$. The low energy values of these vector masses depend upon the scale at which they were generated.  For the new generation to ameliorate the tuning in the Higgs sector, these vector masses must also be at the weak scale (see section \ref{Sec:TheModel}). This requirement leads to a   ``$\mu$-problem" in this model. It is conceivable that the physics responsible for creating these masses is tied to the generation of the $\mu$ parameter of the MSSM. Since $\mu$ is often tied to SUSY breaking, we will assume that these vector masses are also generated at this scale $M_S$. With vector masses $\sim 200$ GeV at $M_S \sim 10^{9}$ GeV, we find that the weak scale masses for the colored particles are $\sim 300 \text{ GeV}$. Under the same conditions, the electroweak particles receive masses  $\sim 200 \text{ GeV}$. 

The renormalization group analysis shows that it is theoretically possible for the new generation to have reasonably large Yukawa couplings $y_4, \, z_4 \subset \l 0.8, 1.1 \r$ to the Higgs sector without losing perturbative unification. In this context, with SUSY breaking $M_S \lessapprox 10^9$ GeV, this scenario can yield soft scalar masses $\sim 400$ GeV, while remaining consistent with experimental bounds on gaugino masses. This parameter space also supports vector masses $\sim 200$ GeV for the  colored components of the new generation. In the next section, we examine the mass spectrum of this model after electroweak symmetry breaking. 

\section{The Mass Spectrum}
\label{Sec:Spectrum}
Upon electroweak symmetry breaking, the $SU(2)$ doublet $\Qn$ splits to yield an up-type quark $\Qn^u$ and a down-type quark $\Qn^d$. In addition to the vector mass, these quarks also receive mass from the Higgs vev. The  mass matrices for these quarks can be expressed as 
\begin{equation}
\begin{pmatrix}
\Qn^u & \UnBar 
\end{pmatrix}
\begin{pmatrix}
\muQ & y_4 \, v \\
0 & \muU \\
\end{pmatrix}
\begin{pmatrix}
\QnBar^u \\ \Un
\end{pmatrix}
\qquad
\text{and}
\qquad
\begin{pmatrix}
\Qn^d & \DnBar 
\end{pmatrix}
\begin{pmatrix}
\muQ & 0 \\
z_4 \, v & \muD \\
\end{pmatrix}
\begin{pmatrix}
\QnBar^d \\ \Dn
\end{pmatrix}
\label{Eqn:UpDownMasses}
\end{equation}
where the superscripts $u$ and $d$ denote the up and down components of the  doublet. 

The mass eigenstates are obtained by bidiagonalizing the above mass matrices. In the limit $\muQ \approxeq \muU  \approxeq \muD = \munew$ and $\munew \gtrapprox y_4 \, v, z_4 \, v$, the eigenvalues simplify to 
\begin{equation}
M_u \approxeq \begin{pmatrix}
\munew - \frac{y_4 \, v}{2} + \frac{\l y_4 \, v\r^2}{8 \, \munew} \\
\munew + \frac{y_4 \, v}{2} + \frac{\l y_4 \, v\r^2}{8 \, \munew}
\end{pmatrix},
\qquad
M_d \approxeq \begin{pmatrix}
\munew - \frac{z_4 \, v}{2} + \frac{\l z_4 \, v\r^2}{8 \, \munew} \\
\munew + \frac{z_4 \, v}{2} + \frac{\l z_4 \, v\r^2}{8 \, \munew}
\end{pmatrix}
\label{Eqn:EigenMasses}
\end{equation}
Electroweak symmetry breaking splits the spectrum into two mass eigenstates, one with mass  $\sim \munew - \frac{v}{2}$ and the other of mass  $\sim \munew + \frac{v}{2}$. In the next section, we study the experimental constraints on these new particles from direct collider searches and precision electroweak observables.


\section{Experimental Constraints}
\label{Sec:Constraints}

The collider signatures of the new quarks in superpotential \eqref{Eqn:NewYukawas}  depend upon its decay channels to the standard model. If the new generation mixes with the standard model, these quarks will decay to the standard model through $W$ or $Z$ emission. These decay channels are constrained by CDF,  which imposes a lower bound  $\gtrapprox 256$ GeV on the mass of a new down-type quark \cite{CDFSearch, Kribs:2007nz}. The bound on a new up-type quark depends upon its branching fraction for decays to $W$s.  If this branching fraction is 100 \%, CDF imposes a lower bound $\gtrapprox 311$ GeV \cite{CDFUpSearch}. However, in this model, this branching fraction can be significantly smaller since it depends upon the unknown mixing angle between the new generation and the standard model. In particular, when both $y_4$ and $z_4$ are non-zero, it is possible for the mass of the lightest down-type quark to be smaller than that of the lightest up-type quark. But, their mass difference could nevertheless be smaller than the $W$ mass (see equation \ref{Eqn:UpDownMasses}). In this case, the up-type quark can dominantly decay to the down-type quark and soft standard model final states as long as it is heavier than the down-type quark by a few GeV. The CDF search does not limit this scenario. Consequently, we will take a lower bound of $\gtrapprox 256$ GeV on the mass of the new quarks.

It is also possible that the new generation is endowed with a parity that forbids it from mixing with the standard model. We discuss this phenomenology in section \ref{Sec:Parity} where we show that as a result of the new parity, the new quarks are meta-stable on collider time scales.  The lifetimes are somewhat model dependent, but can easily be $\sim 10^6$ s even for a 500 GeV quark. The strongest constraint on this decay mode comes from a CDF search for meta-stable CHAMPS \cite{Aaltonen:2009kea}. This study constrains the mass of a meta stable up-quark to be $\gtrapprox 350 - 400$ GeV. The bound on a meta-stable down quark would be weaker by $\sim 30 - 40$ GeV \cite{Acosta:2002ju}. In this model, it is possible for the meta-stable, lightest colored particle to be a down-type quark (see equation \ref{Eqn:UpDownMasses}). Consequently, in this scenario, the lower bound on the quark mass would be $\gtrapprox 300 - 350$ GeV.

Irrespective of the mixing between the new generation and the standard model families, this model requires the new generation to couple to the Higgs. This Yukawa interaction contributes to the precision electroweak parameters $S$, $T$ and $U$ and will impose a restriction on the vector masses.  Using the mass matrices \eqref{Eqn:UpDownMasses}, we computed the corrections to $S$, $T$ and $U$ in this model using \cite{Lavoura:1992np}. Taking $y_4, \, z_4 = 0.9$ and the vector masses $\muQ = \muU = \muD = \munew$, we get 

\begin{eqnarray}
\delta T & = & 0.17 \l\frac{\text{300 GeV}}{\munew}\r^2\nonumber \\
\delta S & = & 0.06 \l\frac{\text{300 GeV}}{\munew}\r^2 \nonumber \\
\delta U & = & 0.004 \l\frac{\text{300 GeV}}{\munew}\r^2
\label{Eqn:PEWCvalues}
\end{eqnarray}

These contributions are within the 68\% confidence limits on these electroweak parameters \cite{PDG, Kribs:2007nz}. These corrections decouple as $\sim \munew^{-2}$ with the vector mass $\munew$ and are quickly suppressed beyond $\munew > 300$ GeV. 

The corrections discussed in Eqn.~\eqref{Eqn:PEWCvalues} were computed for the new fermion sector. Their scalar partners also contribute to these quantities. However, since the scalars are heavier than the fermions, their contributions are more suppressed. Using \cite{Drees:1990dx} to estimate these corrections, we find that with vector masses $\munew \gtrapprox 300$ GeV and soft scalar masses $\widetilde{m}^2 \gtrapprox \l 350 \text{ GeV} \r^2$, the net electroweak contributions from the new sector are within the 95\% confidence limits on the electroweak observables. This constraint was also obtained using $y_4, \, z_4 = 0.9$. The electroweak corrections are sensitive to the squares of the Yukawas $y_4$ and $z_4$. Hence, if either of these Yukawas are a bit small, their contributions are rapidly suppressed and these constraints can be satisfied with even smaller soft scalar masses.

The primary electroweak precision constraint is due to the $T$ parameter, $\delta T \lesssim 0.2$.  Note that this seems to be the origin of the difference between our conclusions and those of \cite{Babu:2008ge}.  While we find that this constraint can be satisfied for $\mu \gtrapprox 300$ GeV, they require $\mu \sim$ TeV.  They seem to have used a formula for $\delta T$ that is a factor of 4 larger than ours.  We would find their result if we had a symmetric mass matrix instead of the asymmetric matrix in Equation \eqref{Eqn:UpDownMasses}.  Further, this factor of roughly 4 between a symmetric and an asymmetric matrix is confirmed by \cite{Maekawa:1995ha}.  Taking the vector mass to be a TeV requires also raising the scalar soft masses to around a TeV in order to get a significant contribution to the Higgs quartic, thus recreating the fine-tuning necessary to get the correct Higgs vev.

In order to accommodate these experimental constraints, we will assume that the vector masses $\munew \gtrapprox 325$ GeV for the rest of the paper.  This forces all new fermions to be heavier than $\gtrapprox 256$ GeV (see equation \eqref{Eqn:EigenMasses}), in agreement with the CDF bound on rapidly decaying particles.  Note that colored fermions at 256 GeV would still be in conflict with collider searches if these particles are stable on collider timescales. Consequently, $\munew \gtrapprox 325$ GeV satisfies all constraints only when the new generation mixes with the standard model. In the absence of this mixing, the collider bound forces the vector mass $\munew \gtrapprox 380$ GeV so that the lightest colored particle is heavier than $\sim 300$ GeV. 

The leptonic spectrum is also split by electroweak symmetry breaking (see equation  \eqref{Eqn:UpDownMasses}).  However, if these leptonic Yukawas are small (see section \ref{Sec:TheModel}), this part of the particle spectrum will be relatively unaffected by electroweak symmetry breaking. Consequently, we expect the new leptons to be roughly around their vector masses $\muL$ and $\muE$. The most stringent constraints on these particles come from LEP, which imposes a lower bound $\muL, \, \muE  > 101$ GeV  \cite{PDG, Kribs:2007nz}.

\section{The Higgs Mass}
\label{Sec:HiggsMass}
In this section, we calculate the corrections to the Higgs mass from the new generation. We restrict the parameter space of the model to that allowed by current experimental bounds, namely, $\munew \gtrapprox 300$ GeV and the condition that the Yukawa couplings remain perturbative up to the GUT scale.  

The correction to the Higgs mass was computed using the one loop effective potential method.  The mass matrices in equation \eqref{Eqn:UpDownMasses} for the quarks $\Qn, \, \QnBar, \, \Un, \, \UnBar, \, \Dn, \, \DnBar$ were bi-diagonalized. For the scalars, diagonal SUSY breaking soft scalar masses  $\widetilde{m}_{{\Qn}}^2, \, \widetilde{m}_{{\QnBar}}^2, \, \widetilde{m}_{{\Un}}^2, \, \widetilde{m}_{{\UnBar}}^2, \, \widetilde{m}_{{\Dn}}^2$ and  $\widetilde{m}_{{\DnBar}}^2$  were added to the supersymmetric masses in equation \eqref{Eqn:UpDownMasses}.  We also added $A$ terms $\sim -300$ GeV to the scalar mass matrix. $A$ terms of this size at the weak scale can be naturally realized from renormalization group effects even if they are negligible at the high scale (see section \ref{Sec:RGE}). The resultant scalar mass matrix was  diagonalized. The mass eigenstates obtained from this procedure were used to calculate the one loop effective potential. This potential was  used to obtain the Higgs mass in the decoupling limit. 

\begin{figure}
\begin{center}
\includegraphics[width = 4.1 in]{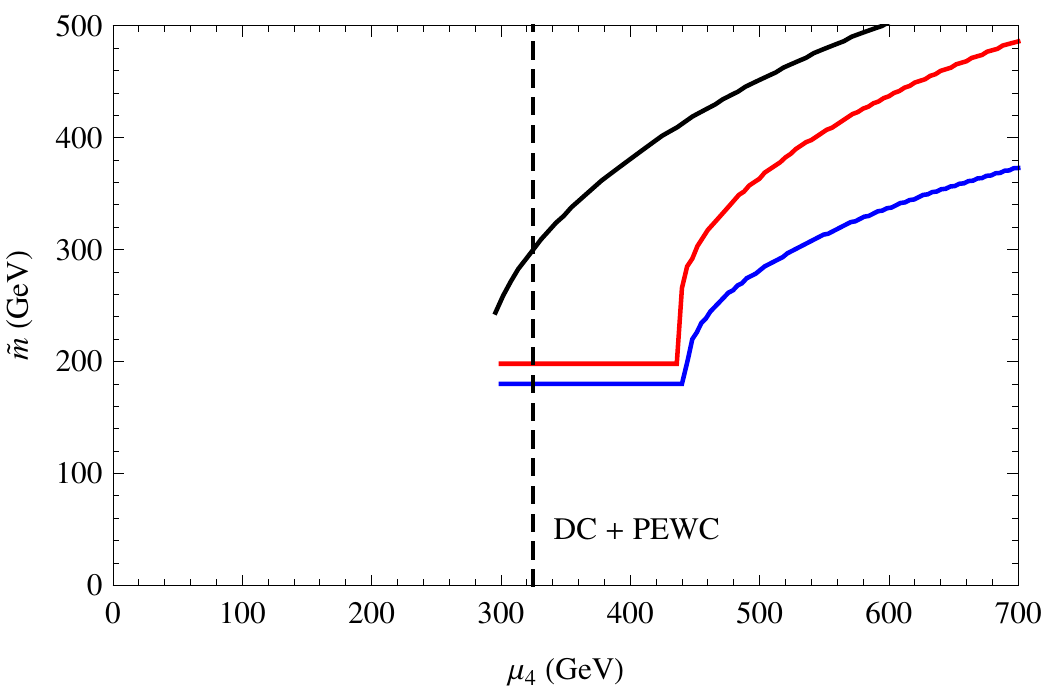}
\caption{ \label{Fig:mqmuplot} (Color online) The parameter space in the $\mu_4$-$\widetilde{m}$ (vector mass - soft mass) plane.  The solid curves are the minimal values of $\widetilde{m}$ that raise the Higgs above 114 GeV and raise all squarks above the bounds from colliders.  From top to bottom they are $(\tan \beta, A) = $ $(5, 300 \, \GeV)$ (black), $(5, 350 \, \GeV)$ (red), $(10, 300 \, \GeV)$ (blue), where $A$ is the unified A-term value.  The lower two curves become flat where they are determined by the lower bound on the top squark mass, not the Higgs mass.  Here we have taken $y_4 = z_4 = 0.9$ and the decoupling limit.  The dashed line is the lower limit on $\mu_4$ from direct collider constraints on the fermions and precision electroweak constraints (DC + PEWC).}
\end{center}
\end{figure}
\begin{figure}
\begin{center}
\includegraphics[width = 3.5 in]{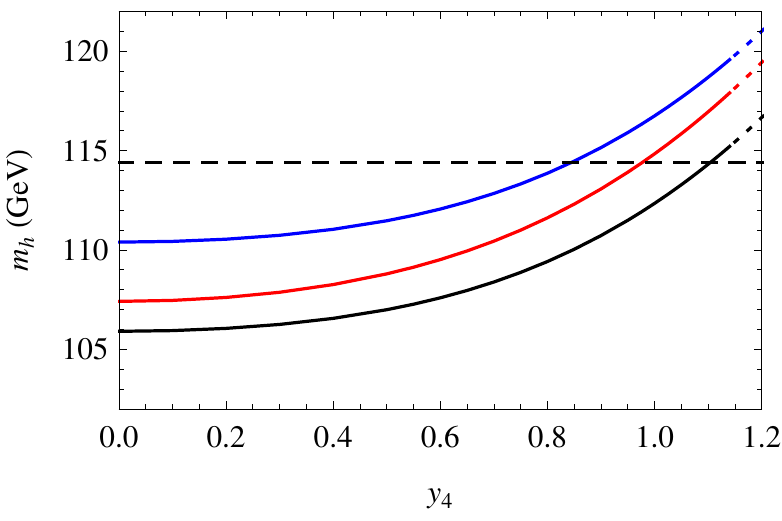}
\caption{ \label{Fig:Higgsplot} (Color online) The Higgs mass versus $y_4$ for $z_4 = 0$.  From bottom to top they are $(\tan \beta, A) = $ $(5, 300 \, \GeV)$ (black), $(5, 350 \, \GeV)$ (red), $(10, 300 \, \GeV)$ (blue), where $A$ is the unified A-term value.  Here we have taken a unified vector mass $\mu_4 = \muQ = \muU = \muD = 320 ~ \GeV$, a unified soft mass $\widetilde{m}^2 =  \widetilde{m}_{Q}^2 = \widetilde{m}_{{U}}^2 =  \widetilde{m}_{{D}}^2 = \left( 350 ~\GeV \right)^2$, and the decoupling limit.  Each curve is dotted where it hits the perturbativity limit from Fig.~\ref{Fig:yzplot}.  The dashed line is the lower limit on the Higgs mass.}
\end{center}
\end{figure}

The above computations were performed numerically and their results are summarized in Figures \ref{Fig:yzplot}, \ref{Fig:mqmuplot} and \ref{Fig:Higgsplot}. Figure \ref{Fig:yzplot} displays the parameter space in the $y_4 \, - \,  z_4$ plane which solves the little hierarchy problem while remaining consistent with perturbative unification.  This parameter space contains regions where $z_4$ is negligible and the Higgs mass corrections arise from $y_4$. This suggests that this framework can solve the little hierarchy problem with just the coupling $y_4  \, \Qn \, \Un \, H_u$. This coupling requires the addition of just a vector-like antisymmetric $SU(5)$ tensor $10$.  The little hierarchy problem can be solved in this manner, although, the allowed range of $y_4$ is smaller. This is because the gauge couplings run less without the contributions from the vector-like $\l 5_4, \, \bar{5}_4\r$ and hence they compete less against the Yukawa contributions in equation \eqref{Eqn:BetaFunctions}. This causes the Yukawas to become non-perturbative more rapidly, restricting $y_4 \lessapprox 1.05$.

In addition to producing a Higgs mass in excess of the LEP bound, our model must satisfy the experimental constraints on the masses of the stop and the fourth generation states. In Figure \ref{Fig:mqmuplot} we plot the allowed region of the vector mass - soft mass parameter space for various values of $\tan \beta$ and a unified A-term $A = A_{y} = A_{z}$. The vertical dashed line indicates the lower bound on the vector mass discussed in the previous section. LEP searches place a lower bound on the lightest stop mass of 96 GeV \cite{PDG}; for given ($\tan \beta$, $A$) this places a lower bound on the soft mass parameter $\tilde{m}$.  This lower bound appears as the horizontal sections of the two lower curves in Figure \ref{Fig:mqmuplot}. For values of $\tilde{m}$ above this bound the parameter space is constrained by the Higgs mass bound, producing the rising parts of the curves. Given these lower bounds on $\mu_4$ and $\tilde{m}$, the fourth generation squarks are automatically above their experimental bounds \cite{Aaltonen:2009kea}. If the soft scalar masses of the first two generations are similar to that of the third, then our assumption of small $A$ terms for the first two generations ensures that those squarks are heavier than the lightest stop. We see that the Higgs mass in this model is larger than the LEP bound even with soft scalar masses $\widetilde{m} \sim 300$ GeV. By contrast, in the MSSM, the squarks have to be $\gtrapprox 1$ TeV to overcome this bound.

In Figure \ref{Fig:Higgsplot}, we study the Higgs mass as a function of $y_4$ for $z_4 = 0$, again with various choices of the parameters $\tan{\beta}$ and $A$.  In this model, Higgs masses up to approximately 120 GeV are possible. This illustrates the possibility of solving the little hierarchy problem adding only the couplings to the vector-like $({\bf 10},\bf{\bar{10}})$ multiplets.


\section{Collider Phenomenology}
\label{Sec:Pheno}
The Yukawa interactions between the new quarks and the Higgs are the biggest source of corrections to the Higgs mass. These quarks have to be light in order to significantly affect the Higgs mass (see sections \ref{Sec:TheModel} and \ref{Sec:HiggsMass}). For moderate $\tan \beta \, \l \approx 5 \r$ and soft SUSY-breaking masses $\widetilde{m} < 400$ GeV, we expect the vector mass $\munew$ of the new quarks to be $\lessapprox 550$ GeV (see Figure \ref{Fig:mqmuplot}) so that the Higgs mass is larger than 114 GeV. This vector mass implies the existence of two sets of quarks,  one with mass $\lessapprox 450$ GeV and the other with mass $\lessapprox 650$ GeV (see equation \eqref{Eqn:UpDownMasses}). Of course, the vector mass can be higher than $550$ GeV if $\tan \beta$ is larger. For example, with $\tan \beta = 10$, the vector mass can be as large as $700$ GeV (giving rise to one $\sim 600$ GeV quark) and still make the Higgs sufficiently heavy. The model  thus predicts an abundance of light colored particles at the LHC: standard model squarks with masses $\lessapprox 500$ GeV, two new quarks and their superpartners, with at least one of the quarks being lighter than $\sim 600$ GeV. 

The existence of the leptonic sector of the new generation is not required to explain the Higgs mass (see section \ref{Sec:TheModel}). However, we expect this new sector to exist in order to preserve gauge coupling unification. The model by itself does not place direct bounds on the vector mass of these leptons. But, if the vector mass of the new generation has a common, unified origin,  the leptons will be significantly lighter than the quarks due to renormalization (see section \ref{Sec:RGE}).  Hence, we expect leptons with masses $\ll 500$ GeV in this model.

In this section, we discuss the observability of these new particles at the LHC. The phenomenology of this model depends upon the mixing between the new generation and the standard model. In the following, we first discuss the case where the new generation mixes with the standard model and then examine the situation where this mixing is forbidden by a parity. We conclude this section with a discussion of the Higgs phenomenology in this model. 

\subsection{Mixing with the Standard Model}
\label{Sec:NoParity}
Mixing between the new generation and the standard model leads to rapid (on collider time scales) decays of the new generation to the standard model. The new quarks will decay through the production of $W$s. The subsequent leptonic decays of the $W$ is a standard, low background way to search for these quarks at the LHC.  With 100 $\text{fb}^{-1}$ of data, the LHC reach for these new quarks is at least $\sim 700$ GeV \cite{ATLASTDR, AguilarSaavedra:2009es}. This reach should cover most of the expected mass range of the new quarks. 

The phenomenology of the new lepton sector with vector masses is different from that of a new chiral lepton sector. In particular, the neutrinos of this sector will have large $\sim 100$ GeV masses. This changes the collider signatures of this new lepton sector since the heavy neutrinos can also decay to the standard model. In conventional searches for fourth generation chiral leptons, the fourth generation neutrinos are assumed to be massless  \cite{ATLASTDR}, preventing them from decaying to the standard model. On the other hand, a new heavy neutrino can decay to produce lepton-rich signals, discriminating it more from standard model background. The LHC reach for this novel lepton sector needs further study \cite{ATLASTDR, Buckley:2009kv, AguilarSaavedra:2009ik}. 

\subsection{Discrete Parity}
\label{Sec:Parity}

The new generation might respect a discrete parity that forbids it from mixing with the standard model ({\it e.g.}~superpotential \eqref{Eqn:NewYukawas}). In this case, the superpotential \eqref{Eqn:NewYukawas} does not allow the lightest new quark to decay. Naively, this model would appear to be ruled out by stringent constraints on long-lived colored relics. However, it is possible for these quarks to decay to their leptonic counterparts (which are typically lighter than the quarks due to renormalization effects) and the standard model through baryon-number violating operators. A natural source of such operators can be found in supersymmetric GUT theories \cite{Arvanitaki:2008hq}. 

For example, if the superpotential \eqref{Eqn:NewYukawas} is embedded into a $SU(5)$ GUT, the term $y_4 \Qn \Un H_u$ can emerge from the $SU(5)$ invariant operator $y_4 10_4 10_4  H_u$. Using the Higgs triplet $H_u^{\l3\r}$, this GUT operator yields  $y_4 \Un \En H_u^{(3)}$. Integrating out the heavy Higgs triplets and using the familiar interactions between the triplets and the standard model, we get the dimension five operator in the superpotential

\begin{equation}
W \supset y_4 y_{b} \frac{\Un \En U_3 D_3}{\Mgut} 
\label{Eqn:Dim5Decay}
\end{equation}
which leads to the decay of the quark  to its leptonic counterpart and the standard model. The decay lifetime is

\begin{equation}
\tau \sim 3 \times 10^8 \text{ s }  \l\frac{1}{y_4}\r^2 \l\frac{2 \times 10^{-2}}{y_b}\r^2 \l\frac{\Mgut}{2 \times 10^{16} \text{ GeV}}\r^2  \l\frac{200 \text{ GeV}}{\delta M}\r^3 
\label{Eqn:LifeTime}
\end{equation} 
where $y_b$ is the bottom Yukawa and $\delta M$ is  the phase space (equal to the mass difference between the colored and leptonic components) available for the decay. 

Similar decay operators can also be generated from embedding the coupling $z_4 \QnBar \DnBar H_u$ into a GUT. However, these operators will also be suppressed by  $y_b$ since the terms in  the superpotential  \eqref{Eqn:NewYukawas} connect the new generation to $H_u$ and not $H_d$. Since a dimension 5 decay through the Higgs triplet must involve couplings to both $H_u$ and $H_d$,  the standard model enters this operator through its interactions with $H_d$. Consequently, these operators are at least suppressed by the bottom Yukawa coupling $y_b$. 

The lifetime in equation \eqref{Eqn:LifeTime} should be regarded as an upper bound for the quarks in a GUT theory, since we expect the Higgs triplet to couple to the light particles in most GUTs. But, it is possible for the quarks to decay more rapidly than \eqref{Eqn:LifeTime}. For example, new physics at the GUT scale can lead to faster decays with lifetimes $\sim 1000$ s \cite{Arvanitaki:2008hq}. 

The bounds on the lifetime of long lived colored particles is sensitive to their abundance. This abundance is uncertain due to non-perturbative processes that occur during the QCD phase transition \cite{Kang:2006yd, Arvanitaki:2005fa}. Under the assumptions of \cite{Kang:2006yd}, decays with lifetimes $\lessapprox 10^{14}$ s are unconstrained by cosmology. But, with the caveats in  \cite{Arvanitaki:2008hq, Arvanitaki:2005fa}, the abundance could be larger than the estimates of \cite{Kang:2006yd}. In this case, it is possible for these decays to have cosmological implications. For example, these decays may help explain the primordial lithium problems  \cite{Arvanitaki:2008hq, Jedamzik2008}. 

Regardless of their cosmological impact, long lived colored particles give rise to striking signals in colliders \cite{Arvanitaki:2005nq, Kraan:2005ji}. The colored particles, upon production, will travel through the detector where they will lose energy due to electromagnetic and hadronic interactions, giving rise to charged tracks in the detector. A fraction of these particles will stop in the detector and eventually decay. Since these decays are uncorrelated with the beam, they can be distinguished from most backgrounds. The LHC reach for such meta-stable quarks is at least 1 TeV \cite{Fairbairn:2006gg}.  

The discrete parity also leads to interesting phenomenology in the lepton sector, since it stabilizes the lightest leptonic particle. If the lighest particle is the new neutrino, it is a natural candidate for WIMP dark matter. This neutrino by itself couples too strongly to the $Z$ and is in conflict with bounds from dark matter direct detection experiments. However, this problem can be solved if this neutrino mixes with a standard model singlet $S_4$ through the term $x_4 \Ln S_4 H_u$ \cite{TuckerSmith:2001hy}. The singlet $S_4$ and the  Yukawa coupling $x_4$ could emerge naturally in a $SO(10)$ GUT from the $SO(10)$ invariant Yukawa $16_4 16_4 10_h$. 


\subsection{Higgs Phenomenology}
\label{Sec:HiggsProduction}
Gluon fusion is the dominant production channel for the Higgs at the LHC. The new quarks contribute to this  channel and can enhance the production cross-section. This enhancement is smaller than the case of a chiral fourth generation \cite{Kribs:2007nz} since the additional contributions are suppressed by the vector mass $\munew$. The enhancement should roughly scale as $\sim \l y_4^2 + z_4^2\r\l\frac{v}{\munew}\r^2$ and is $\sim \frac{1}{3} \l\frac{300 \text{ GeV}}{\munew}\r^2$ for $ y_4, \, z_4 \sim \OrderOne$. 

The mass spectrum may also permit the decay of the Higgs to the new leptons. It is possible for this decay channel to be the dominant decay mode of the Higgs since we expect the new generation to have $\OrderOne$ Yukawas to the Higgs. The subsequent decay of the new leptons to the standard model might offer a new way to discover the Higgs and the new leptons. This possibility merits further study. 

\section{Conclusions}
A new vector-like generation with $\OrderOne$ Yukawa couplings has interesting phenomenological consequences. Owing to the vector mass, collider and precision electroweak constraints are more easily avoided. This is unlike the case of a chiral fourth generation where these constraints rapidly force the theory to become non-perturbative  \cite{Kribs:2007nz}. Low energy non-perturbativity can be avoided in such models at the cost of drastically reduced tree-level contributions to the Higgs mass, further accentuating the little hierarchy problem \cite{Kribs:2007nz}. A vector mass, on the other hand, easily avoids experimental constraints and simultaneously allows a solution to the little hierarchy problem. Furthermore such a solution is predictive.  The new generation can significantly ameliorate the tuning of the Higgs vev if it has a mass between $\sim 300 - 500$ GeV, along with soft scalar masses  between $\sim 300 - 400$ GeV. The spectrum contains many light, colored particles which would be well within the LHC reach.

Our model motivates consideration of several novel collider signatures.  A vector-like generation can give rise to unique decay channels for the new leptons and neutrinos.  The new generation also changes the collider phenomenology of the Higgs sector. In addition to increasing the Higgs production cross section, the new generation allows for non-standard Higgs decay channels. It is also possible for this framework to contain long-lived quarks. These quarks give rise to striking signatures at colliders and may help solve the primordial lithium problems.  The LHC phenomenology of this new sector deserves further consideration.

In this paper, we discussed the effects of a new vector-like generation on the Higgs mass. In addition to its effects on Higgs physics, a new vector-like generation could also have other phenomenological uses. For example, it may help explain the hierarchy in the fermion mass spectrum  \cite{Graham:2009gr, Dobrescu:2008sz}. It could also arise in string constructions where the number of chiral generations emerges as a result of a mismatch between the number of left-handed and right-handed chiral fields. In such a construction, it may also be possible to address the new $\mu$ problem raised by this scenario. It is also possible for these new $\OrderOne$ Yukawa couplings to modify the electroweak phase transition and stimulate electroweak baryogenesis. In this case, there might be additional signatures of this model, for example, gravitational wave signals that may be observable in upcoming experiments  \cite{Dimopoulos:2008sv,Dimopoulos:2007cj, LISAPPA}.  The presence of such a fourth generation would clearly have important implications for UV physics beyond just this model.


{\it Note added:} While this paper was in the final stages of preparation, \cite{Martin:2009} appeared which has some overlap with this work.

\section*{Acknowledgments}
We would like to thank Chris Beem, Csaba Csaki, Savas Dimopoulos, Dan Green, David E. Kaplan, Roni Harnik, Stephen Martin and Stuart Raby for useful discussions and the Dalitz Institute at Oxford for hospitality.  S.R. was supported by the DOE Office of Nuclear Physics under grant DE-FG02-94ER40818.  S.R. is also supported by NSF grant PHY-0600465.  A.I. was supported by the Natural Sciences and Engineering Research Council of Canada.

\end{document}